\documentclass [12pt]{article}
\usepackage {epsfig}
\usepackage{graphicx}

\begin{document}

\begin{center}
{\Large {\bf On the Energy Spectrum of Cosmogenic Neutrons}}

\vskip 0.5cm
{\bf
A.S.~Malgin
}
~~~

{\it  Institute for Nuclear Research RAS, Moscow, Russia}
\end{center}

\noindent E-mail: Malgin@lngs.infn.it

\begin{abstract}

The processes of the generation of cosmogenic neutrons (cg-neutrons) underground are considered. The neutrons produced by cosmic-ray muons 
in their interactions with matter are called cosmogenic. Deep-inelastic ${\pi A}$-collisions of pions in muon-induced hadronic showers are 
mainly their source at energies above 30 MeV. The characteristics of the energy spectrum for the generation of cg-neutrons have been determined 
by invoking the additive quark model of deep-inelastic soft processes and the mechanism for the interactions of high-energy nucleons in a nucleus. 
The three-component shape of the spectrum is explained, and the energy of the ``knee'' in the spectrum has been found to depend on the mass number ${A}$. 
The peculiarities of deep-inelastic ${\pi A}$-scattering lead to the conclusion that the spectrum of cg-neutrons steepens sharply at energies above 
1 GeV. The calculated quantitative characteristics of the spectrum are compared with those obtained in measurements.
\end{abstract}


\vskip 0.5cm
\noindent
{\bf
A full version was published in Journal of Experimental 
and Theoretical Physics, 2017, Vol. 125, No. 5, pp. 728-740.
}
\vskip 1.5cm

\section{Introduction}
\label{intro}
The characteristics of muon-induced cosmogenic (cg) neutrons in various materials deep underground have been actively 
studied in the last 15 years. These neutrons are among the main sources of the background in the numerous low-background underground
experiments that are now underway, with which hopes to detect effects beyond the Standard Model are associated: hypothetical dark 
matter particle interactions, neutrinoless double ${\beta}$-decay, and other rare phenomena.

Today it can be stated that one of the main characteristics of the cg-neutron flux, the yield ${Y_n}$, has been studied 
quite well. A relationship of the yield ${Y_n}$ to the energy losses of ultrarelativistic muons and the mass number of 
matter ${A}$ has been established on the basis of experimental data \cite{1}. An analytical expression that
allows the yield ${Y_n}$ and, consequently, the flux and production rate of cg-neutrons to be calculated with a sufficient 
accuracy for any ${A}$  at depths greater than
100 m w.e., where the mean energy of the muon flux ${\overline E_{\mu} >}$ 40 GeV, has been derived.

This cannot be said about another important characteristic, the energy spectrum for the generation (the ``in-source'' spectrum) 
of cg-neutrons ${F^s(T_n)}$, where ${T_n}$ is the neutron kinetic energy. The spectrum ${F^s(T_n)}$ determines the 
spectrum ${F^{is}(T_n)}$ of isolated neutrons (neutrons at the boundary of the semi-infinite layer of material where 
they are generated), the penetrating power of cg-neutrons, and the set of possible effects produced by them in detectors, shielding materials,
and soil.

\section{The generation of cosmogenic neutrons}

The existing general views of the mechanism for the generation of cg-neutrons by ultrarelativistic muons were shaped within the theory of muon-induced hadronic ($h$) and electromagnetic ($em$) showers in matter by invoking the ideas of intranuclear nucleon cascades (INCs) and neutron photoproduction \cite{2},\cite{3}.

The basic tenets of these views are:

(a) cg-neutrons at depths greater than 100 m w.e. are generated in any material mainly in showers; the neutrons from ${{\mu}^-A}$-captures dominating at depths up
to ${\sim}$ 100 m w.e. and the photoneutrons produced by a muon through virtual photons constitute a minor fraction in the total number of cg-neutrons;

(b) in any material $h$-showers make a major contribution to the production of neutrons, despite the much smaller generation cross section of $h$-showers than that of $em$-showers, which is explained by the greater multiplicity of neutrons ${{\nu}_n}$ in $h$-showers;

(c) the overwhelming majority of $h$-shower neutrons are produced by the ${{\pi}N}$-interactions of shower charged pions with nucleons in nuclei; the neutrons generated in INCs ($cas$-neutrons), including the ${{\pi}N}$-scattering recoil neutrons, lie in the entire energy range of cg-neutrons; at energies ${T_n >}$ 30 MeV their mean  ${\overline T_n^{cas}}$  is $\sim$ 150 MeV; ``evaporative'' neutrons ($ev$-neutrons) escaping from remnant nuclei whose number is 
approximately twice that of $cas$-neutrons and whose energy is ${T_n^{ev} <}$ 30 MeV appear in the last INC phase; a small number of 
$h$-shower neutrons are produced in ${\pi}^-A$-captures and $em$-subshowers initiated by ${\pi}^0$-decays, these neutrons have energies up to 30 MeV;

(d) the overwhelming majority of $em$-shower neutrons are produced by photons through the giant dipole resonance (GDR), i.e., they are evaporative and have an energy up to 30 MeV; the neutrons from the direct photoeffect (neutron photospallation), the $\gamma$-absorption by the deuterium ${np}$-pair in a nucleus, and the neutrons of $h$-subshowers appearing in emshowers as a result of multiple pion photoproduction can have energies above 30 MeV, but the contribution of the listed channels to the total yield of cg-neutrons is minor;

(e) the spectrum ${F^s(T_n)}$ is formed by $ev$- and $cas$-neutrons; the number of $ev$-neutrons ${N_n^{ev}}$ generated in $em$- and $h$-showers exceeds considerably the number of $cas$-neutrons ${N_n^{cas}}$; at ${T_n >}$ 30 MeV the spectrum ${F^s(T_n)}$ has two segments, flat and steeper, the most common representation of the spectral shape is the dependence ${1/T^{\alpha}}$ with different exponents ${\alpha}$.

In the absence of a general understanding of the cg-neutron generation mechanism, the above tenets do not give an answer to the following practical questions: (1) whether the dependence ${T^{-\alpha}}$  corresponds to the cg-neutron generation processes and the exponent ${\alpha}$ of what quantity can characterize the spectrum ${F^s(T_n)}$  at ${T_n >}$  30 MeV before and after the knee; (2) at what energy ${T_n^{ch}}$ the spectrum changes and what processes cause the knee in the spectrum; (3) whether there can be a ``cutoff'' in the spectrum and, if so, as a result of what process and at what energy; (4) what fraction in the total number are accounted for by the evaporative neutrons; (5) how the quantities listed in items (1) - (4) depend on the energy ${\overline E_{\mu}}$ and the mass number ${A}$ of matter.

Establishing the shape of the spectrum ${F^s(T_n)}$ from experimental data and answering the above questions are complicated by the limited number of measurements
and their significant errors. In the time of experimental investigation since 1954 \cite{4} the energy characteristics of cg-neutrons generated by muons with various energies ${\overline E_{\mu}}$  in various materials have been measured in six experiments \cite{5},\cite{6}, \cite{7}, \cite{8}, \cite{9}, \cite{10}. The spectrum of isolated neutrons ${F^{is}(T_n)}$ at energies below 90 MeV was obtained in \cite{5},\cite{6}. The neutron generation spectrum ${F^s(T_n)}$ was determined in the last four papers: up to 80 MeV in \cite{7}, \cite{10} and up to 400 MeV in \cite{8}, \cite{9}.

The calculations performed in recent years for various materials using the up-to-date FLUKA and Geant4 software packages \cite{11}, \cite{12}, \cite{13}, \cite{14}, \cite{15}, \cite{16} (Fig. 1), being in significant disagreement between themselves in spectral shape (and in yield ${Y_n}$), do not allow both the characteristics of the spectrum ${F^s(T_n)}$ for a specific material and their relation to the energy ${\overline E_{\mu}}$  and ${A}$ to be determined computationally.

\begin{figure}[!t] 
 \begin{center}
  \includegraphics[width=\linewidth]{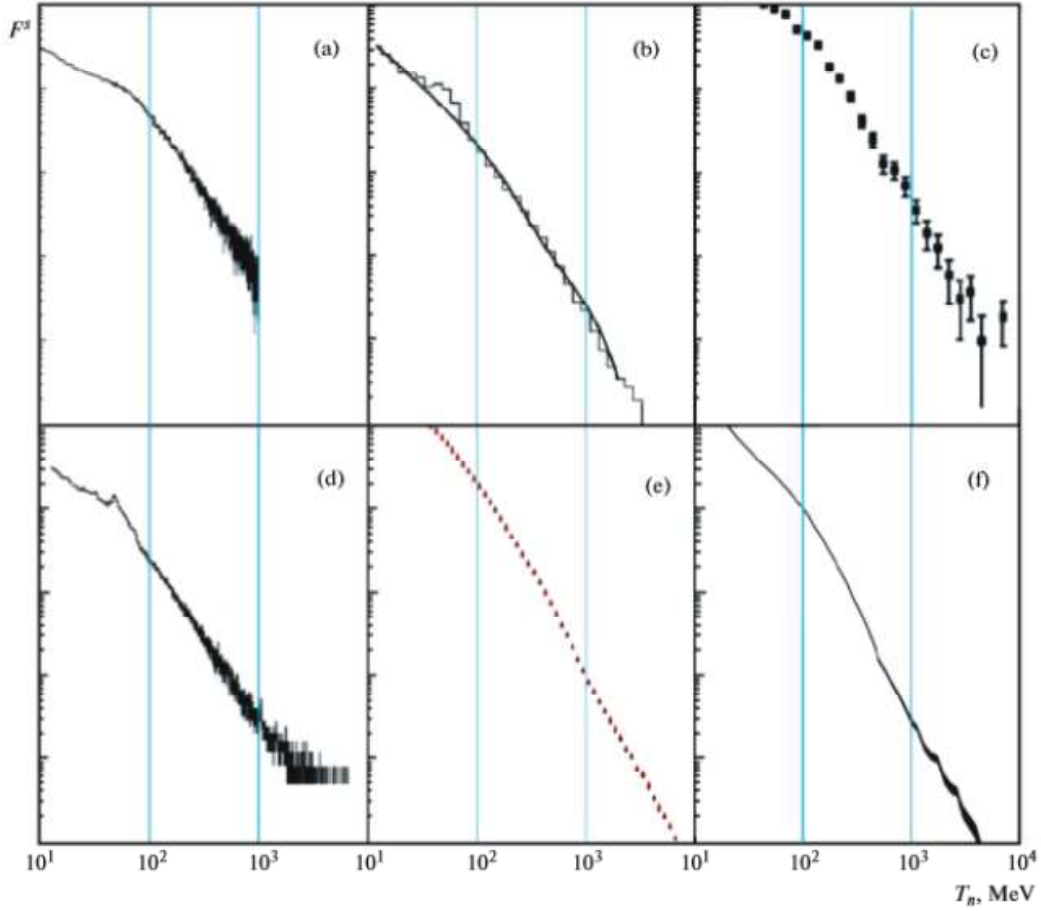}
 \end{center} 
  \caption{Energy spectra $F^s(T_n)$ for the generation of neutrons by muons of fixed energy $E_{\mu}$ calculated by the Monte Carlo method for various materials: (a) LS, FLUKA, $E_{\mu}$= 285 GeV \cite{12}; (b) LS, FLUKA and Geant4 (histogram), $E_{\mu}$ = 280 GeV \cite{11};
(c) Pb, FLUKA, $E_{\mu}$ = 300 GeV \cite{14}; (d) LS, Geant4, $E_{\mu}$ = 280 GeV \cite{13}; (e) soil, Geant4, $E_{\mu}$ = 275 GeV \cite{15}; (f) Pb, Geant4, $E_{\mu}$ = 260 GeV \cite{16}.} 
  \label{1fig} 
 \end{figure}

\section{The production of neutrons in h-showers}

The most penetrating neutrons forming the spectrum ${F^s(T_n)}$  at ${T_n >}$ 30 MeV are produced in $h$-showers. These showers are initiated by deep-inelastic ${\mu}A$-scattering ${\mu}A \to k{\pi} + mN + (A - m) +{\mu}'$ ($k$ and $m$ are the multiplicities of the produced pions and nucleons, ${m {\leq} A}$) and develop in matter due to the multiplication of pions in deep-inelastic ${\pi}A$-collisions: ${\pi}A \to k{\pi} +mN + (A - m)$. In both ${\mu}A$- and ${\pi}A$-reactions the multiple production of pions is the result of an inelastic collision of the incident particle with one of the intranuclear
nucleons. The multiple generation of pions occurs mainly in deep-inelastic soft processes, which make a major contribution to the cross section for the
deep-inelastic interaction of hadrons. These processes are characterized by small transverse momenta of generated pions $p_{\perp} <$  1 GeV/$c$, in contrast to the hard ones with $p_{\perp} >$ 1 GeV/$c$.

Let us consider the production of neutrons in $h$-showers in terms of the quark model of strong interactions.
In the additive quark model the elastic $qq$-scattering
of two quasi-free valence (structural) quarks
(current quarks interact in hard processes) underlies
the soft multiple processes. This process can be
depicted by a quark diagram (Fig. 2). In this representation
of ${\pi}N$-interaction the antiquark of the incident
pion ${\overline q_{\pi}}$ is elastically scattered by the quark $q_N$  of the target nucleon $N$. Both quarks acquire momenta sufficient for their escape from the confinements of the interacting hadrons. The breaking of the gluon bond of
the scattered antiquark ${\overline q_{\pi}}$ to the spectator quark $q_{\pi}$ gives rise to virtual $q{\overline q}$-pairs with their subsequent
recombination into a jet of fragmentation pions. The
number of recombined $q{\overline q}$-pairs and, consequently,
pions is determined by the energy of the antiquark ${\overline q}_{\pi}$.
There is also a leading pion with an energy of about
1/2 of the initial one produced by the spectator quark
$q_{\pi}$  and the pair antiquark ${\overline q}$ among the fragmentation
pions. The escape of the quark $q_N$ from the nucleon
also gives rise to $q{\overline q}$-pairs and their recombination into
pionization hadrons. The spectator diquark $q_Nq_N$  is
hadronized (``decolorized'') and turns into a recoil
nucleon $N_r$ ($r$-nucleon).

 \begin{figure}[!t] 
  \begin{center}
  \includegraphics[width=\linewidth]{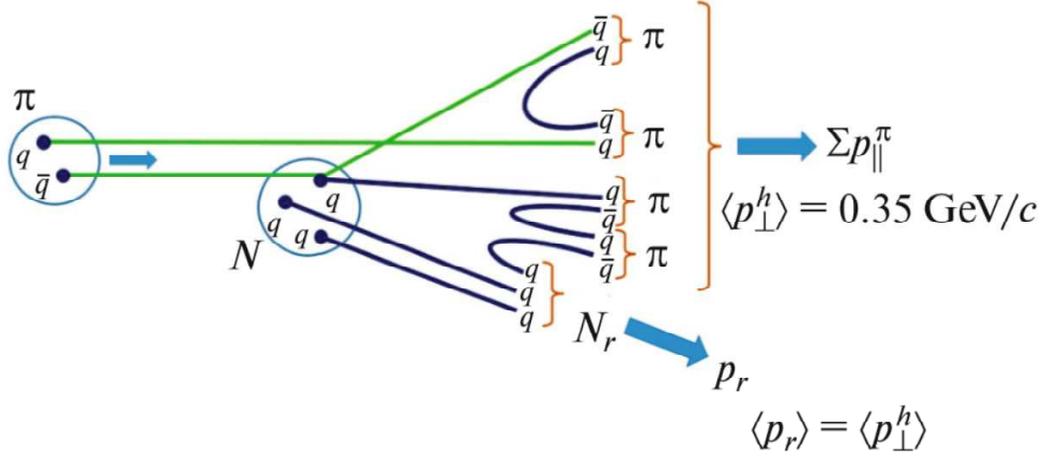}  
  \end{center}
  \caption{Quark diagram for the collision of valence quarks in deep-inelastic ${\pi}N$-scattering to produce a recoil nucleon $N_r$ in the laboratory frame.} 
  \label{2fig}
 \end{figure}

It is well known from experiments in cosmic rays and on accelerators that during the collision of particles with ultrarelativistic energies the pions (hadrons) in the
forming jets have an interaction-energy-independent
mean transverse momentum ${\langle} p_{\perp}^h {\rangle} {\approx}$ 0.35 GeV/$c$. In the QCD theory of strong interactions this quantity corresponds to the scale of gluon vacuum fluctuations with a characteristic nucleon size ${\sim}$ 1 Fm ($10^{-13}$ cm), which is the nucleon confinement size. When gaining an additional momentum (in our case, from ${\overline q_{\pi}}$), the quark $q_N$ can reach the confinement boundary (nucleon periphery). Here, the dipole gluon field binding the escaping quark to the diquark $q_Nq_N$  is compressed into a string
that breaks if the quark has a sufficient momentum. Multiple pion production is a consequence of the breaking.

The gluon string tension is determined by the strong coupling constant. The constant and, consequently, the string tension on the nucleon periphery grow sharply with increasing distance to the escaping quark $q_N$. The string breaks, on average, at the same tension at a distance $r_q$ ${\sim}$ 1 Fm irrespective of the escaping-quark momentum. As a result, the $r$-nucleon formed from the diquark $q_Nq_N$  gains a momentum $p_r$ directed along the motion of the escaping quark in the
nucleus whose mean value must correspond to the
string tension at a confinement size ${\sim}$ 1 Fm, i.e., ${\langle} p_r {\rangle}$ ${\approx}$ 0.35 GeV/$c$, which leads to the equality ${\langle} p_r {\rangle}$ ${\approx}$ ${\langle} p_{\perp}^h {\rangle}$.
Hence it follows that the mean energy transferred to the $r$-neutron is independent of the ${\pi}N$-interaction energy and mass number $A$. The subsequent fate of the
$r$-neutron is determined by its energy and location in the nucleus. The shower pions produced in ${\pi}N$-collisions escape from the nucleus mostly without any collisions, because their mean free path in nuclear matter, ${\overline \lambda}_{{\pi}N}$ ${\sim}$  5 Fm, exceeds the radii of almost all nuclei.

\section{The energy spectrum of recoil neutrons $F({\mathbf E}_r)$}

In accordance with the expression $pc = {\beta}E$ ($E = m_Nc^2 + T_N$), the $r$-nucleon, gaining the momentum ${\langle} p_r {\rangle}$ = 0.35 GeV/$c$, acquires the energy ${\overline {\mathbf E}_r}$ = 63 MeV in addition to the energy of its Fermi motion in the nucleus $T_F$ ${\approx}$ 30 MeV. The recoil nucleon can remain in the nucleus, leave the nucleus without any collision, or initiate an INC in an elastic $NN$-collision. During the
INC the $r$-nucleon energy is fragmented; some of the $cas$-nucleons (including the neutrons) can escape from the nucleus. In an elastic $NN$-collision the
energy is divided between the nucleons, on average, equally unless such a proportion is forbidden by Pauli blocking. Therefore, at the mean energy transfer ${\overline {\mathbf E}_r}$ = 63 MeV the $cas$-neutrons escaping from the nucleus
have mostly energies in the ``evaporative'' range approximately up to 30 MeV. Consequently, the neutron energy spectrum at $T_n$ ${\geq}$ 30 MeV is formed by the $r$-neutrons escaping from nuclei without any collision.

Since the $r$-neutron is a product of elastic $q_{\pi}q_N$-scattering, the spectrum $F({\mathbf E_r})$ of $q_N$-quark energy transfers to the $r$-neutron must be characterized by the dependence $F({\mathbf E_r})$ ${\propto}$ 1/${\mathbf E_r^2}$ corresponding to the elastic scattering of two quasi-free point charges. The ${\delta}$-electrons
produced by charged particles during their ionization stopping in matter have the same spectrum. The $r$-neutron that is capable of producing nuclear particles ($p, {\alpha},n, t$) in inelastic $nA$-reactions after its escape from the nucleus corresponds to the definition of a ${\delta}$-particle.
The $r$-neutron in the nucleus must gain the minimum energy
${\mathbf E}_r^{min}$ = 
${\varepsilon}_b + {\varepsilon}_{th}$ = 15 MeV to be transformed into a ${\delta}$-neutron, because its binding energy in the nucleus is ${\varepsilon}_b$ ${\approx}$  8 MeV and the threshold of inelastic $nA$-reactions is ${\varepsilon}_{th}$ ${\approx}$ 7 MeV \cite{17}.

The existence of minimum energy transfer ${\mathbf E}_r^{min}$ at a given mean energy ${\overline {\mathbf E}}_r$ of the spectrum of a known shape, 
$F({\mathbf E_r})$ = const${\mathbf E_r^{-2}}$  , points to the existence of a
limiting $r$-neutron energy ${\mathbf E}_r^{max}$. Its value can be found using the definition of the mean:

\begin{equation}
{\overline {\mathbf E}}_r = \int\limits_{{\mathbf E}_r^{min}}^{ {\mathbf E}_r^{max} }{\mathbf E}_rF({\mathbf E}_r)d{\mathbf E}_r/\int\limits_{{\mathbf E}_r^{min}}^{{\mathbf E}_r^{max}}F({\mathbf E}_r)d{\mathbf E}_r\\
=\frac{{\mathbf E}_r^{min}{\mathbf E}_r^{max}ln({\mathbf E}_r^{max}/{\mathbf E}_r^{min})}{{\mathbf E}_r^{max}-{\mathbf E}_r^{min}}
\end{equation}

At ${\mathbf E}_r^{max}$ ${\gg}$ ${\mathbf E}_r^{min}$  this expression takes the form $\overline \mathbf E_r$ ${\approx}$ ${\mathbf E_r^{min}}$ ln( ${\mathbf E_r^{max}}$ / ${\mathbf E_r^{min}}$ ); substituting $\overline \mathbf E_r$  = 63 MeV and
$\mathbf E_r^{min}$ = 15 MeV here yields ${\mathbf E_r^{max}}$ ${\approx}$ 1 GeV. ${\mathbf E_r^{max}}$ is the maximum energy transfer to the $r$-nucleon in deepinelastic soft ${\pi}N$-scattering that is determined by the tension energy of the gluon string between the valence quark  $q_N$  and the diquark $q_Nq_N$ before its breaking.

Note that the ``transverse'' energy of 1 GeV corresponding to the momentum $p_{\perp}$ = 2 GeV/$c$ is at the boundary of the region where the hard processes
involving current quarks begin to contribute to the total deep-inelastic scattering cross section.

As follows from Eq. (1), at fixed ${\overline {\mathbf E}}_r$ the energy ${\mathbf E_r^{max}}$ depends strongly on ${\mathbf E_r^{min}}$. Despite being physically justified, its choice contains certain arbitrariness that admits a deviation of about 50\% from the chosen value. Thus, at ${\mathbf E_r^{min}}$ = 11.5 MeV and ${\mathbf E}_r^{min}$ = 18.5 MeV for ${\mathbf E_r^{max}}$  we obtain 2.8 and 0.6 GeV, respectively. Assuming that the range of transverse energies below 1 GeV includes the overwhelming majority of soft processes, below we adopt ${\mathbf E}_r^{max}$ = 1 GeV.

\section{The spectrum of ${\delta}$-neutrons}

The spectrum of ${\delta}$-neutrons $F(T_{\delta})$ is formed by the $r$-neutrons
(Fig. 2) that escaped from the nuclei without any collisions ($T_{\delta}$ is the neutron energy outside the nucleus). The energy spectrum of $r$-neutrons before their escape from the nucleus $F(T_r)$ is specified by the spectrum of energy transfers $F({\mathbf E_r})$ including the Fermi energy $T_F$: $F(T_r) = F({\mathbf E_r} +T_F)$. When escaping freely from the nucleus, the $r$-neutron loses the energy $T_F$ and the binding energy ${\varepsilon}_b$ ${\approx}$ 8 MeV; consequently, $T_{\delta}^{min}$ = ${\mathbf E_r^{min}}$ $-$ ${\varepsilon}_b$ ${\approx}$ 7 MeV and $T_{\delta}^{max}$ = ${\mathbf E_r^{max}}$ $-$ ${\varepsilon}_b$ ${\approx}$ $10^3$ MeV. The effect of limiting nuclear fragmentation, i.e., the number of secondary slow particles and the mean energy of ${\delta}$-nucleons do not depend on the primary particle energy if it exceeds ${\sim}$ 5 GeV \cite{18}, \cite{19}], detected in photoemulsion experiments points to a ``cutoff'' of the ${\delta}$-neutron spectrum.

The existence of maximum energy $T_{\delta}^{max}$ based on the model of deep-inelastic soft ${\pi}N$-scattering does not mean a complete cutoff of the cg-neutron spectrum. There are processes that give rise to ${\delta}$-neutrons with an energy above 1 GeV. These primarily include the elastic ${\pi}N$-scattering in the range of pion energies $E_{\pi} >$ 10 GeV, where neutrons with an energy $T_{\delta} >$ 1 GeV can be produced. The elastic ${\pi}N$-scattering cross section,
${\sigma}_{\pi N}^{el}$ ${\approx}$ 4 mb, accounts for no more than 20\% of the inelastic one, ${\sigma}_{\pi N}^{in}$ ${\approx}$ 20 mb. Other processes, such as the neutron stripping by real and virtual high-energy photons and deep-inelastic hard ${\pi}N$-scattering, make a contribution less than 2\% of the elastic scattering
contribution. Thus, by the cutoff of the cg-neutron spectrum at an energy of ${\sim}$ 1 GeV we mean its sharp steepening.

The probability of the escape of an $r$-neutron from
nuclei without any collision is determined by the place
of its appearance in the nucleus (the ${\pi}N$-interaction
point) and the mean free path in the nuclear matter ${\overline {\lambda}_r}$.
The distribution of ${\pi}N$-collisions in the nucleus does
not change with shower pion energy but is related to
the mass number $A$. The cross section for deep-inelastic
pion scattering by a free nucleon in the energy range
5 $-$ 100 GeV is virtually constant: ${\sigma}_{{\pi}N}$ ${\approx}$ 20 mb \cite{20}. The effective ${\pi}N$-scattering cross section in the nucleus,
${\sigma}_{{\pi}N}^{eff}$, is smaller due to the nucleon shadowing: ${\sigma}_{{\pi}N}^{eff}$ = ${\sigma}_{{\pi}N}A^{\alpha - 1}$, where ${\alpha}$ = 0.75 is the shadowing parameter \cite{21}. Thus, the mean free path of a high-energy pion in nuclear matter, ${\overline {\lambda}_{{\pi}N}}$, is expressed by the dependence ${\overline {\lambda}_{{\pi}N}}$ = $({\sigma}_{{\pi}N}^{eff}{\overline n_{\chi}})^{-1}$ cm, where ${\overline n_{\chi}}$ is the average nucleon concentration on the pion path ${\chi}$. The pion, on average, crosses a nucleus of radius $R$  along the chord ${\overline {\chi}}$ = $4R/3$ = $[4(1.3A^{1/3})/3]$ Fm = 1.73$A^{1/3}$ Fm located at a distance of about $3R/4$ from the nuclear center. The distribution of nucleon concentration along the nuclear radius is described by the expression

\begin{equation}
n(R) = n_0(1 + exp \frac{R-R_0}{\Delta}),
\end{equation}

where $n_0$ = const is the nucleon concentration in the core of the nucleus, which is related to the mean nucleon concentration in the nucleus, ${\overline n}$, 
by the relation $n_0$ = 1.9 ${\overline n}$; $R_0$ = 1.08$A^{1/3}$ Fm is the distance from
the nuclear center to the region with the number density $n_0/2$; $\Delta \approx$ 0.55 Fm is the rate of decrease in concentration $n$ on the nucleus periphery. The concentration ${\overline n_{\chi}}$ is related to $n_0$  and the mean concentration of nucleons in the nucleus, ${\overline n}$ = $10^{38}$ cm$^{-3}$: 
${\overline n_{\chi}}$ $\approx$ 0.4 $n_0$  $\approx$ 0.76 ${\overline n}$. Substituting ${\overline n_{\chi}}$ and ${\sigma}_{\pi N}^{eff}$ into the expression for ${\overline {\lambda}_{{\pi}N}}$, we obtain ${\overline {\lambda}_{{\pi}N}}$ $\approx$ 6.58 $A^{1/4}$ Fm. The integral probability of a ${\pi}N$-collision, $P_{{\pi}N}$, on a path length ${\overline {\chi}}$ is determined by the ratio 
${\overline {\chi}} / \overline {\lambda}_{{\pi}N}$:
\[
P_{{\pi}N} = 1 - exp (-\overline \chi / \overline {\lambda}_{{\pi}N}).\nonumber
\]
The quantity $\overline \chi / 
\overline {\lambda}_{{\pi}N}$ is weakly related to $A$ ( $\overline \chi / \overline {\lambda}_{{\pi}N}$  = 0.263 $A^{1/12}$); it can be assumed to be a constant, $\overline \chi / \overline {\lambda}_{{\pi}N}$ $\approx$ 0.368 at $A^{1/12}$ $\approx$ 1.4. In that case, $P_{\pi N}$ is also a constant:
 $P_{\pi N}$ = 0.308. 

We will consider the median length $l_m$ in the distribution
of ${\pi}N$-collisions along the chord $\overline {\chi}$ to be the effective
distance from the entrance point of the pion into the nucleus to its collision with the nucleon. It is determined from the equation 0.5 $P_{{\pi}N}$ 
= 1 $-$ $exp(-l_m/\overline {\lambda}_{{\pi}N})$:

\begin{equation}
l_m = - \overline {\lambda}_{{\pi}N} {\mathrm {ln}}(1 - 0.5P_{{\pi}N}) = 1.10A^{1/4} \mathrm {Fm}.
\end{equation}

Consequently, the $r$-neutron appears at the distance ${\overline l}_r$ = $\overline {\chi}$ $-$ $l_m$ = 1.32 $A^{1/4}$ Fm along the chord $\overline {\chi}$ before its escape from the nucleus. When the neutron momentum is directed in a cone along the motion of the incident pion, the condition for the escape of neutrons from the nucleus without any collision is that the mean free path of the $r$-neutron in the nucleus, ${\overline \lambda}_r$ =
$({\overline \sigma}_{nN} {\overline n}_{\chi})^{-1}$, exceeds the length ${\overline l}_r$: ${\overline \lambda}_r \geq {\overline l}_r$ . Equating the $r$-neutron mean free path ${\overline l}_r$ to the length ${\overline {\chi}}  - l_m$ is some approximation, because the recoil neutron is
deflected from the trajectory of the pion entering the nucleus under the action of the gluon string tension and its intranuclear motion. The cross section ${\overline {\sigma}_{nN}}(T_n)$ is an averaged cross section for neutron scattering by nucleons in the nucleus,

\begin{equation}
{\overline \sigma}_{nN} = \frac{(A-Z){\sigma}_{nn} + Z{\sigma}_{np}}{A}, 
\end{equation}

where ${\sigma}_{nn}$ and ${\sigma}_{np}$ are the effective cross sections for neutron scattering by a free neutron and proton, $Z$ is the number of protons in the nucleus. The dependences of the cross sections ${\sigma}_{nn}$ and ${\sigma}_{np}$ on energy $T_n$ are presented in Fig. 3. In the range of energies $T_n$  from 30 to $\sim$ 200 MeV ${\sigma}_{np}$  is larger than ${\sigma}_{nn}$ by a factor of 3 $-$ 1.5. Since $A - Z$ $ \approx Z$ for most nuclei, we can assume that in the neutron energy range 30 $-$ 200 MeV ${\overline \sigma}_{nN}$   is virtually independent of the nucleus composition ${\overline \sigma}_{nN}$ $\approx$ $({\sigma}_{nn} + {\sigma}_{np} )/2$ and ${\overline \sigma}_{nN}$ $\propto$ $ 1/T_n$. At these energies ${overline {\sigma}}_{nN}$ can be approximately calculated from 
the formula ${overline {\sigma}}_{nN}$ ($T_n$) $\approx$ $(6 \times 10^3/T_n)$ mb, where $T_n$ is in megaelectronvolts. 

 \begin{figure}[!t] 
  \centering 
  \includegraphics[width=2.5in]{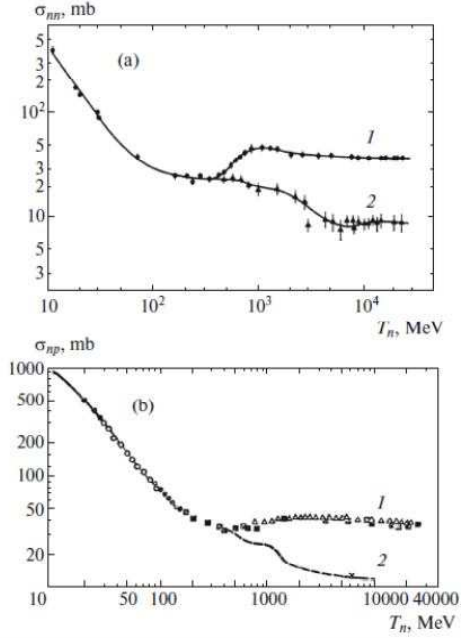} 
  \caption{Total (1) and elastic (2) cross sections for neutron-neutron (a) and neutron-proton (b) scattering} 
  \label{3fig} 
 \end{figure}

The equality ${\overline \lambda}_r = {\overline l}_r$ and the expressions for ${\overline \lambda}_r$ and ${\overline l}_r$  allow us to determine the critical cross section ${\overline {\sigma}_{nN}^{cr}}$ = $100 A^{1/4}$ mb and then the corresponding energy $T_r^{cr}$:
$T_r^{cr}$ = $6 \times 10^3/{\overline \sigma}_{nN}^{cr}$ $\approx$ $60 A^{1/4}$ MeV. This is the threshold value for a free $r$-neutron escape from the nucleus. The energy 
$T_{\delta}^{cr}$ = $T_r^{cr}$ $-$ ($T_F  + {\varepsilon}_b$) $\approx$  ($T_r^{cr} - 40$) MeV in the spectrum of $\delta$-neutrons $F(T_{\delta})$, corresponds to the energy $T_r^{cr}$ in the spectrum $F(T_r)$. Consequently, at $\delta$-neutron energies $T_{\delta} \geq T_r^{cr}(A)$ the spectrum $F(T_{\delta})$
closely follows the shape of the spectrum $F({\mathbf E}_r) \propto$
$1/{\mathbf E}_r^2$ , i.e., $F(T_{\delta}) \propto 1/T_{\delta}^2$.

At $T_r  < T_r^{cr}$ the fraction $P_{\delta}$ of $r$-neutrons freely escaping from the nucleus is determined by the ratio ${\overline {\lambda}_r}$/${\overline l}_r$. Transforming it, we obtain

\begin{equation}
P_{\delta} \propto \frac{{\overline {\lambda}_r}} 
                      {{\overline l}_r} 
= \frac{1}{\overline {\sigma}_{nN}(T_r) \overline n_{\chi} \overline l_r(A)} 
\propto T_rA^{-1/4},
\end{equation}

because $\overline n_{\chi}$ is a constant, $\overline l_r \propto A^{1/4}$, and $\overline {\sigma}_{nN}$ $\propto$ $1/T_r$.

Under the action of the factor $P_{\delta}$ the spectrum $F(T_r)$ in the range $\sim$ 45 MeV $< T_r  <T_r^{cr}$ is transformed into the spectrum of $\delta$-neutrons with an energy $T_{\delta}$ from $T_{\delta}^{min}$ to $T_r^{cr}$:

\begin{equation}
F(T_{\delta}) \propto F(T_r)P_{\delta}(T_r) \propto \frac{1}{T_r^2}T_rA^{-1/4} = T_{\delta}^{-1}A^{-1/4}.
\end{equation}

Thus, the spectrum of $\delta$-neutrons in any material spans the range from $T_{\delta}^{min}$ $\sim$ 7 MeV to $T_{\delta}^{max}$ $\sim$ 1000 MeV. The spectrum $F(T_{\delta})$ is $1/T_{\delta}$ in the range from $T_{\delta}^{min}$ to $T_{\delta}^{cr}$  and is described by the law $1/T_{\delta}^2$  in the range $T_{\delta}^{cr}$ - $T_{\delta}^{max}$.

The $A$ dependence of the $\delta$-neutron spectrum $F(T_{\delta}) \propto A^{-1/4}$ is inverse with respect to the $A$ dependence of the knee energy $T_{\delta}^{cr} \propto A^{1/4}$. Consequently, the energy $T_{\delta}^{cr}$ increases with $A$, while the values of the spectrum $F(T_{\delta})$ decrease. This is indicative of a decrease in the fraction of $\delta$-neutrons in the total number of cg-neutrons and an increase in the fraction of $ev$-neutrons. The energy of the knee in the spectrum $F(T_{\delta})$ also decreases with decreasing $A$ and then disappears altogether at $A$ $\sim$ 4, because all $r$-neutrons become $\delta$-neutrons. It should be noted that for nuclei with $A <$ 12 the expression for $T_r^{cr}$ acquires a large error. This is explained by the absence of a core with a constant concentration $n_0$ whose size is specified by Eq. (2).

\section{Measurements of the cg-neutron production spectrum $F^s(T_{\delta})$}

The spectrum $F^s(T_{\delta})$ of cg-neutrons can be obtained in an experiment whose geometry corresponds to the following requirements:

(a) the layer of target material where neutrons are produced must be located near the detector at a fixed distance from it, given that the $\delta$-neutrons are directed
from top to bottom;

(b) the target layer must have a fixed thickness of several mean free paths ${\lambda}_{\pi A}$ but not affecting the shape of the spectrum $F^s(T_{\delta})$;

(c) the presence of a layer of material between the target and the detector must not lead to any deformation of the spectral shape of the neutrons escaping from the target.

In most experiments it is necessary to exclude the detector crossing by a muon and/or a shower, which prevents the energy release of the neutron appearing in the detector from being measured.

The cg-neutron production spectrum $F^s(T_{\delta})$ was obtained in the KARMEN \cite{7} and LVD \cite{8} experiments and the Soudan underground laboratory \cite{10}. In the KARMEN experiment the neutrons were generated by muons and showers mainly in a 7000-ton iron shield surrounding a 65 m$^3$ liquid scintillation
detector. The crossing by a muon of the planes of the active detector shield, which recorded the time and coordinates of the muon-shower event, was a trigger for recording neutrons. The events of iron-shield crossing by a muon containing the detector pulses correlated with the muon were selected for the analysis.
Since the main goal of the KARMEN experiment was to investigate the ${\nu}A$-cross sections in a beam of accelerator neutrinos, the facility was at a small depth
underground. The neutrons from atmospheric muons were studied as a background source in the measurements. The mean energy of the muons passing through the shield corresponded to a depth of $\sim$ 40 m w.e., i.e., it was $\sim$ 15 GeV. Muons of such an energy produce mostly $cv$-neutrons, because the dominant channels of their generation are the ${\mu}^-A$-capture and photoproduction by virtual photons. Nevertheless, recording the temporal, spatial, and energy characteristics of scintillation detector pulses allowed one to establish the energy releases of $\delta$-neutrons escaping from the iron shield and then to determine the shape of the spectrum of 
``visible'' energy releases in the energy range 20 $-$ 80 MeV. The authors of \cite{7} fitted the spectrum of recorded cg-neutron energy releases
$F({\varepsilon}_{vis})$ by the law $exp(- {\varepsilon}_{vis}/{\varepsilon}_0)$, where ${\varepsilon}_0$ $\approx$ 42 MeV (Fig. 4). As follows from the graph, in the range 30 $-$ 80 MeV being discussed 38 and 19 of the 59 experimental data points lie, respectively, above and below
the straight line representing the law $exp(-{\varepsilon}_{vis}/42)$;
two points visually agree with the fit. The discrepancy between the experiment and the dependence $exp(-{\varepsilon}_{vis}/42)$ is most pronounced in the tail of the spectrum: at a measurement uncertainty of about 35\% in the energy range 60 $-$ 80 MeV 17 of the 20 data points lie above the straight line. The exponential with ${\varepsilon}_0$ = 57.8 $\pm$ 9.4 is the best fit to the KARMEN data in the
range 20 $-$ 80 MeV. The significant error in ${\varepsilon}_0$ is related to the determination of count rates from the graph in Fig. 4. Taking this fact into account, we assume below
that ${\varepsilon}_0$ $\sim$ 60 MeV. This value is supported by the result
of the Soudan experiment \cite{10} described below.

 \begin{figure}[!t] 
  \centering 
  \includegraphics[width=\linewidth]{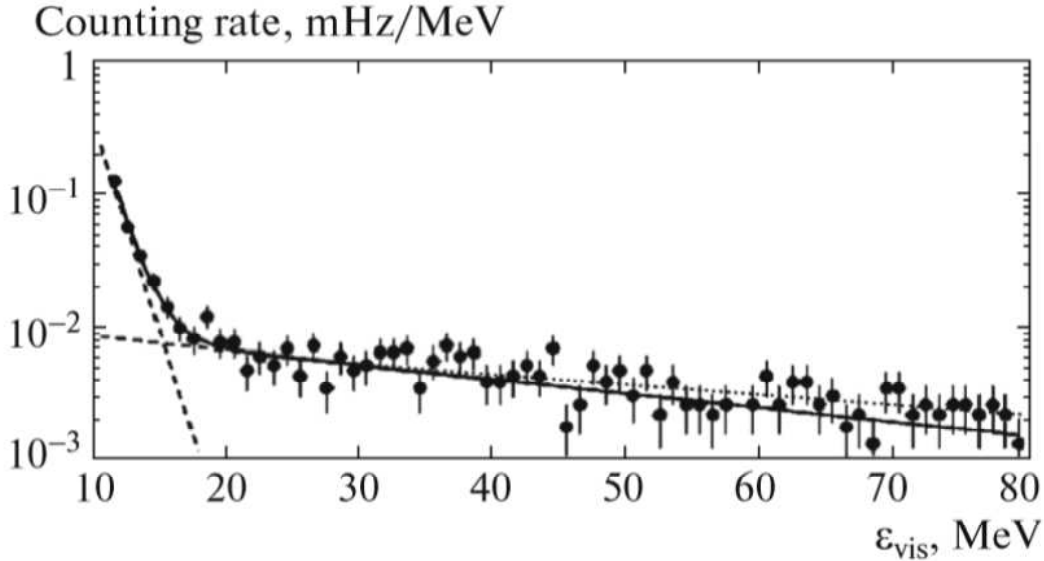} 
  \caption{Spectrum of recorded cg-neutron energies in the KARMEN experiment. The solid line is the fit to the data from \cite{7}. The dashed line is the best fit to the data by the law $exp({\varepsilon}_{vis}/60 MeV)$ in the range 30 $-$ 80 MeV.} 
  \label{4fig} 
 \end{figure}

The relationship between the dependences $F({\varepsilon}_{vis})$ $\propto$ $exp(-{\varepsilon}_{vis}/60)$ and $F({\varepsilon}_{vis}^{\delta})$ is shown in Fig. 5. The curve
$F({\varepsilon}_{vis}^{\delta})$ represents the dependence $N(T_{\delta})$ $\propto$ $T_{\delta}^{-1}$, where the energy $T_{\delta}$ was recalculated to the visible energy releas ${\varepsilon}_{vis}^{\delta}$ using the quenching function $q(T_{\delta})$ = ${\varepsilon}_{vis}^{\delta}/T_{\delta}$ obtained for an organic liquid scintillator (LS) with an approximate composition C$_{10}$H$_{20}$ in \cite{22}. Here, the neutron was assumed to transfer all its energy $T_{\delta}$ to one
recoil proton, i.e., $T_p = T_{\delta}$. In reality, the neutron in LS loses its energy in portions in $np$-scatterings and elastic and inelastic $n$C-collisions. This assumption
affects insignificantly the shape of the curve $F({\varepsilon}_{vis}^{\delta})$. The curves in Fig. 5 were normalized to the number of events at the energy release ${\varepsilon}_{vis}$ = 30 MeV corresponding to $T_{\delta}$ = 47 MeV. The relative positions of the curves
$F({\varepsilon}_{vis})$ and $F({\varepsilon}_{vis}^{\delta})$ show that the KARMEN experimental data in the range 30 $-$ 80 MeV can be fitted by
the law $T_{\delta}^{-1}$.

 \begin{figure}[!t] 
  \centering 
  \includegraphics[width=\linewidth]{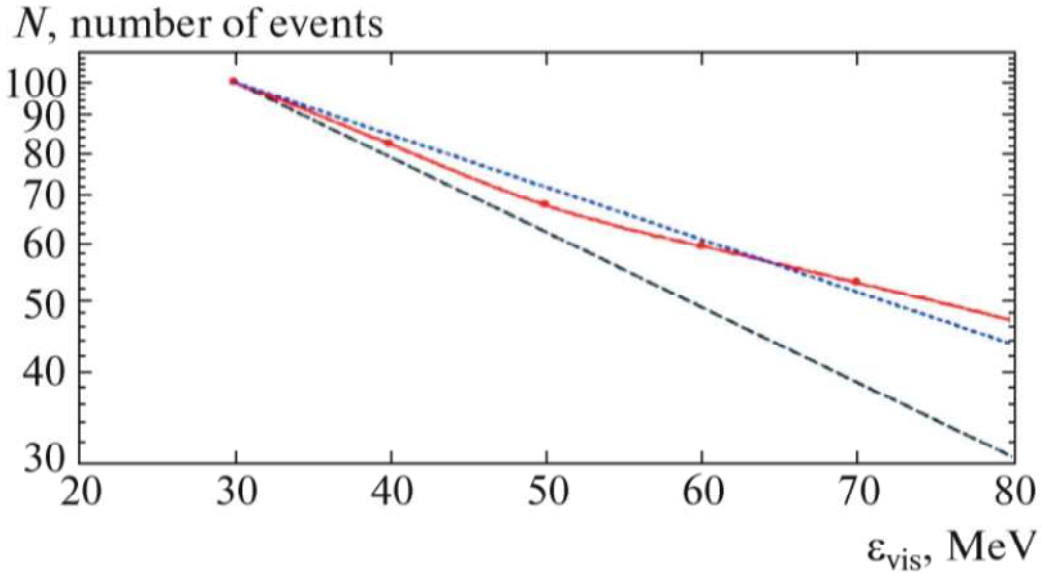} 
  \caption{Spectra of recorded cg-neutron energies ${\varepsilon}_{vis}$ in the range 30 $-$ 80 MeV. The dashed line is the fit to the data 
by the authors of the KARMEN experiment; the dotted line is the fit to the data by the law $exp(-{\varepsilon}_{vis}/60 MeV)$; the solid line 
is the dependence $1/T_n$ corrected for the quenching. The dependences were normalized at ${\varepsilon}_{vis}$ = 30 MeV.} 
  \label{5fig} 
 \end{figure}

The experiment \cite{10} in an underground laboratory at a depth of 2100 m w.e. (${\overline E}_{\mu}$ $\sim$ 210 GeV) at the Soudan mine (USA) was aimed at determining the spectrum $F^s(T_{\delta})$. The measurements were performed with a cylindrical counter ($\oslash$ = 13 cm, $L$ = 100 cm) filled
with 12 $L$ of LS. The range of measured energies ${\varepsilon}_{vis}$ is
from 4 to 60 MeV. The range was limited from above by the electronics capabilities. The interaction of neutrons in LS was separated from the interactions of electrons,
$\gamma$-ray photons, and muons by the PMT pulse shape. The low statistics of 24 events in 655 days of measurements underground did not allow definitive conclusions
about the shape of the spectrum in the range 30 $-$ 60 MeV to be reached. Of interest are the results obtained by the authors during the calibration measurements on the Earth's surface: the spectrum of neutron energies ${\varepsilon}_{vis}$  in the range 30 $-$ 56 MeV, $\propto$ $ exp(-{\varepsilon}_{vis}/60 MeV)$ (Fig. 6), coincided with the corrected KARMEN spectrum (Fig. 4). Thus, the authors of \cite{10} obtained the spectrum $F^s(T_{\delta})$ on the surface. This fact
can be explained by the detection of $\delta$-neutrons generated by extensive air shower (EAS) pions in the roof and walls of the room in which the measurements were made.

 \begin{figure}[!t] 
  \centering 
  \includegraphics[width=\linewidth]{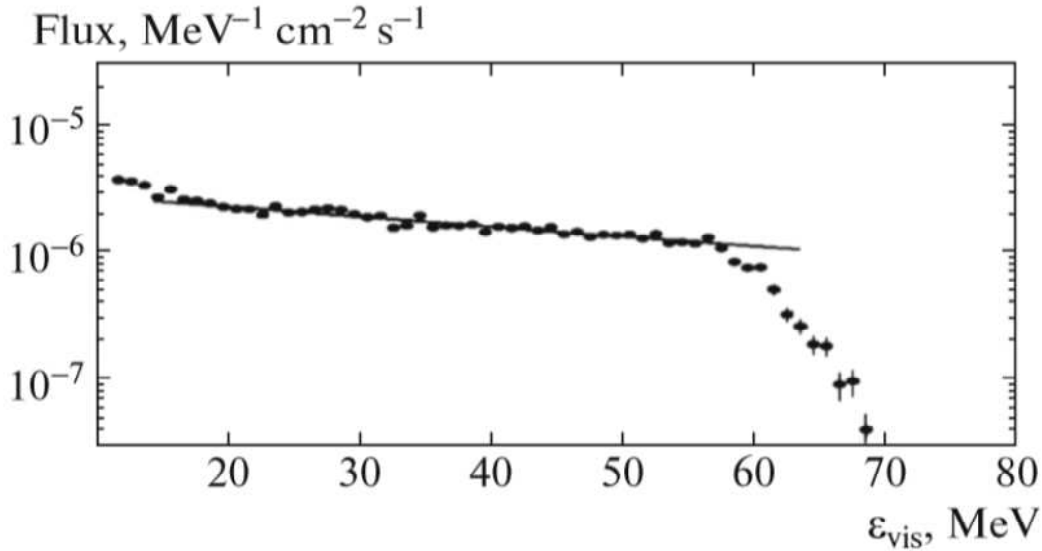} 
  \caption{Observed spectrum of cg-neutron energy releases in the Soudan experiment (dots) \cite{10}. The straight line is the dependence $exp(-{\varepsilon}_{vis}/60 MeV)$.} 
  \label{6fig} 
 \end{figure}

A similar conclusion follows from the measurements of the neutron spectrum in the Gran Sasso National Laboratory at an altitude of 1000 m using a 1.5 m$^3$ scintillator detector \cite{9}. The measured spectrum in the range 30 $-$ 400 MeV with a knee near 90 MeV corresponds to the $\delta$-neutron production spectrum $F^s(T_{\delta})$ obtained underground in the LVD experiment \cite{8}. On the surface neutrons with energies above 30 MeV could also be generated by EAS pions in
the roof of the experimental room. The similarity of the spectra of neutrons at energies above 30 MeV obtained on the surface and underground has a natural explanation: in both cases, they are generated by $h$-shower pions in deep-inelastic ${\pi}A$-scattering, whereby the energy characteristics of the produced $\delta$-neutrons do not depend on the pion energy.

The spectrum $F^s(T_{\delta})$ of cg-neutrons at a depth of 3650 m w.e. was obtained in the LVD experiment (Fig. 7) \cite{8}. The main goal of the experiment is to detect and investigate the neutrino flux from a gravitational stellar core collapse. The LVD characteristics also allow the high-energy muons and their interaction products, including the cg-neutrons, to be studied. The rock overburden above the facility determines a mean energy of the muon flux of 280 $\pm$ 18 GeV. LVD consists
of three identical 6.2 $\times$ 13.8 $\times$ 10.0 m$^3$ towers. Each tower contains 280 rectangular 1.5 m$^3$ counters accommodating 1.2 tons of liquid scintillator each. The counter energy resolution for energy releases above 20 MeV is about 20\%, the time resolution is 1$\mu$s.

 \begin{figure}[!t] 
  \centering 
  \includegraphics[width=\linewidth]{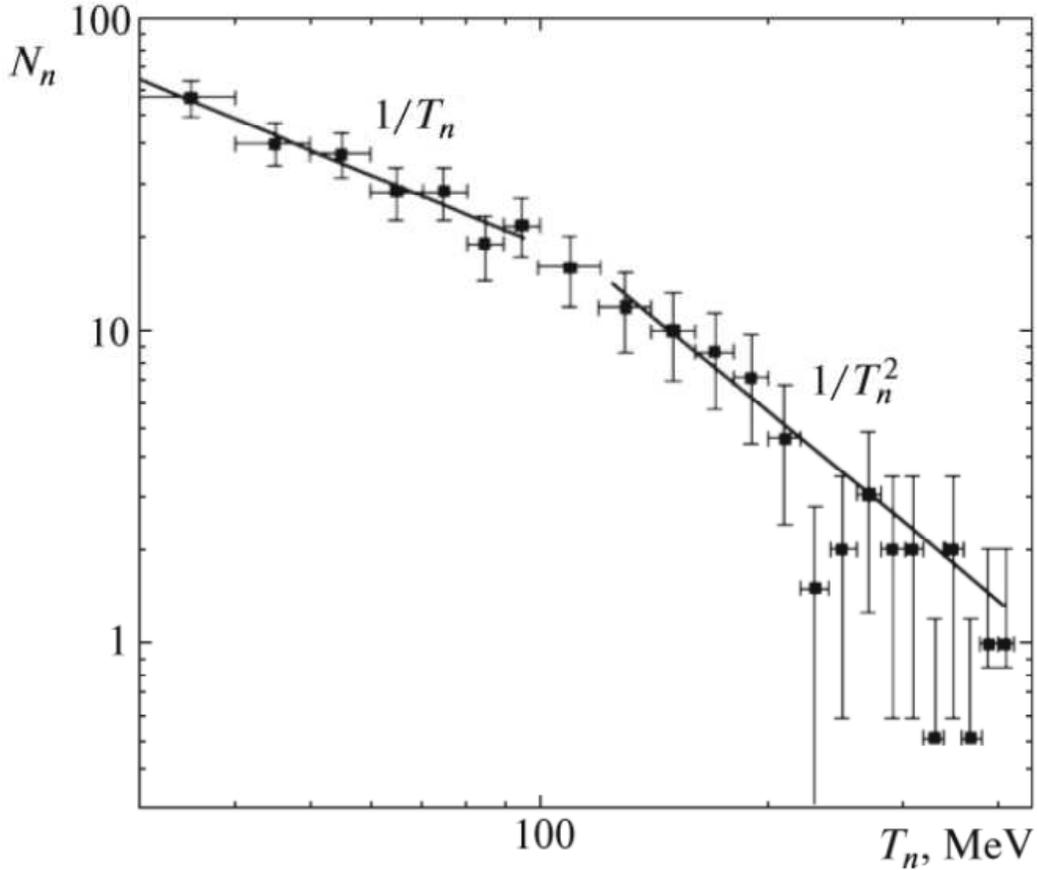} 
  \caption{Energy spectrum of cg-neutrons in the LVD experiment \cite{8}, $N_n$ is the number of neutrons.} 
  \label{7fig} 
 \end{figure}

Neutrons were generated in the material of the facility composed of LS and iron. Iron is contained in the structural elements of the facility forming a regular cellular structure whose cell is a 1.5 m$^3$ scintillation counter. The LS and Fe masses are identical, the LS composition is C$_k$H$_{2k}$, $k$ $\approx$  10. The spectrum $F^s(T_{\delta})$ of neutrons produced by nearly vertical muons in the target column (t-column) of the facility 6.2 m in length, 2.2 m in width, and 7.5 m in height (56 counters with an LS mass of 67 t and the same Fe mass) was reconstructed from the spectrum of ``visible'' neutron energy releases in 60 scintillation counters of the detecting volume (d-volume) with an LS mass of 72 t \cite{8}. The d-volume is separated from the t-column by a 78 g cm$^{-2}$ LS layer and iron layers with a total thickness of 45 g cm$^{-2}$ contained in the separating half-column (28 counters). The half-column, which incorporates
light and heavy materials, efficiently shields the d-volume from the particles of the electromagnetic (${\gamma}, e$) and hadronic ($p, {\pi}^{\pm}$) components of the $em$- and $h$-showers developing in the t-column. At the same time, the separating half-column with this composition and sizes slightly deforms the shape of the spectrum $F^s(T_{\delta})$. This is explained by the following factors. While crossing the half-column, the neutrons with energies $T  >$ 30 MeV interact mainly with C and Fe
nuclei. In elastic $n$C- and $n$Fe-scatterings the neutrons lose less than 10\% of their energy. During an inelastic $nA$-interaction the $\delta$-neutrons with energies above 30 MeV initiate the development of INCs in the nucleus, which leads to fragmentation of the energy $T_{\delta}$ and practically disappearance of the neutrons with an energy above 30 MeV that experienced inelastic $nA$-scattering. Since the cross section for this process, while slightly varying, is essentially independent of the incident neutron energy (Fig. 8), the inelastic scattering affects insignificantly the shape of the spectrum $F^s(T_{\delta})$.

 \begin{figure}[!t] 
  \centering 
  \includegraphics[width=3.0in]{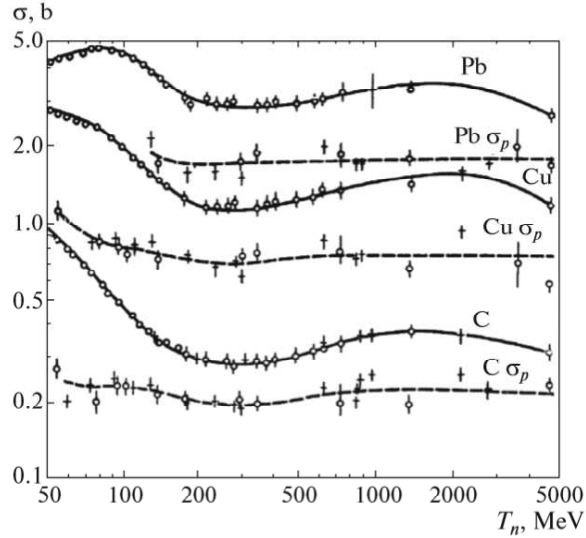} 
  \caption{Total and inelastic (${\sigma}_p$) $nA$-scattering cross sections for C, Cu, and Pb.}
  \label{8fig} 
 \end{figure}

The $np$-scattering, as a result of which the neutron loses, on average, half of its energy, could play a prominent role in the processes under consideration. However,
since the mean free path of $\delta$-neutrons with $T_{\delta} >$ 30 MeV relative to the $np$-scattering in LS, ${\lambda}_{np} >$ 50 cm,
is comparable to the particle mean path length in LS, 75 cm (the mean length of the counter chord) and since the fraction of $np$-collisions in LS is about 20\%,
this process also slightly changes the shape of the $\delta$-neutron spectrum $F^s(T_{\delta})$ at the exit from the separating half-column. Monte Carlo simulations \cite{3}, \cite{14} also suggest a weak influence of material layer with a
thickness up to several mean free paths ${\lambda}_{nA}$ on the spectrum shape for neutrons with energies 10 $-$ 10$^3$ MeV. 

The pulses detected by the d-volume counters in a time interval of 250 ns after a muon event in the target column were attributed to the neutron energy releases.
The sought-for spectrum $F^s(T_{\delta})$ is formed by the neutrons produced in deep-inelastic $\pi$C-, $\pi$Fe-, and ${\pi}p$-interactions whose fractions in the total number of ${\pi}A$-interactions are 48\% ($\pi$C), 37\% ($\pi$Fe), and 15\% (${\pi}p$). In contrast to the spectrum $F^s({\pi}p)$ $\propto$ $ 1/T_{\delta}^{-2}$, the spectra of $\delta$-neutrons in carbon and iron change their shape at 72 and 124 MeV, respectively. Since the fraction
of ${\pi}p$-interactions is small, the measured spectrum $F^s(T_{\delta})$ is formed by the neutrons from $\pi$C- and $\pi$Fe-interactions. As a consequence, the spectrum of $\delta$-neutrons is described by the dependence $T_{\delta}^{-1}$ in the energy range from 30 to $\sim$ 70 MeV and by the law $T_{\delta}^{-2}$ above 120 MeV. The LVD data (Fig. 7) are fitted by the law $T_{\delta}^{-{\alpha}}$ with $\alpha$ = 1.11 $\pm$ 0.30 in the range 30 $-$ 120 MeV and with $\alpha$ = 2.05 $\pm$ 0.14 above 120 MeV \cite{8}. The range 70 $-$ 120 MeV is transitional. Since the $\pi$C- and $\pi$Fe-contributions are comparable, the shape of the sought-for spectrum can be assumed to change at an energy $\sim$ 100 MeV, i.e., $T_{\delta}^{cr}$ (LS + Fe) $\approx$ 100 MeV. Establishing the knee energy $T_{\delta}^{cr}$ allows the characteristics of the spectrum $F^s(T_{\delta})$ in the energy range 30 $-$ 400 MeV corresponding to the measurements to be calculated: the mean energy ${\overline T}_{\delta}$ for the entire range 30 $-$ 400 MeV and the mean energies ${\overline T}_{\delta}^{(1)}$ and ${\overline T}_{\delta}^{(2)}$ before and after the knee.

Our calculations give ${\overline T}_{\delta}^{(1)}$ = 58 MeV (the range 30 $-$ 100 MeV, $F^s(T_{\delta})$  $\propto$ $T_{\delta}^{-1}$) and ${\overline T}_{\delta}^{(2)}$ = 185 MeV (the range
100 $-$ 400 MeV, $F^s(T_{\delta})$ $\propto$ $T_{\delta}^{-2}$). The spectrum-averaged calculated energy in the range 30 - 400 MeV, 
${\overline T}_{\delta}$, is 107 MeV. For the range of measurements 30 $-$ 1000 MeV
the calculated ${\overline T}_{\delta}$ would be 143 MeV. In the experiment
(Fig. 7) for total statistics of 371 neutrons (185 and 186 neutrons before and after the knee, respectively) we obtained ${\overline T}_{\delta}$ = 104 $\pm$ 6 MeV, ${\overline T}_{\delta}^{(1)}$ = 59 $\pm$ 2 MeV, and
${\overline T}_{\delta}^{(2)}$= 174 $\pm$ 14 MeV. It can be seen that the characteristics of the spectrum $F^s(T_{\delta})$ calculated for the energy range of cg-neutrons 30 $-$ 400 MeV are consistent with the measurements. The agreement of the calculated characteristics of the spectrum $F^s(T_{\delta})$ with the results of the available experiments leads to the conclusion that the model for the production of cosmogenic $\delta$-neutrons considered is efficient.

\section{The spectrum of isolated neutrons $F^{is}(T_n)$}

In underground experiments the neutron background component produced by cosmogenic isolated neutrons ($is$-neutrons) is difficult to remove. The neutrons whose appearance is related to the detector neither in time nor spatially are deemed isolated. The neutrons from detector-crossing muons and showers are eliminated by this condition during measurements.

The spectrum of $is$-neutrons underground $F^{is}(T_n)$ is formed at the rock $-$ experimental chamber boundary. Being equiprobably generated in the rock volume,
the cg-neutrons traverse its various thicknesses, reaching the chamber walls. As a result, the neutron production spectrum $F^s(T_n)$ is transformed into the spectrum
$F^{is}(T_n)$. The isolated neutrons in the evaporative region of the spectrum $F^s(T_n)$ at energies $T_n  \leq$ 30 MeV have an isotropic spatial distribution. Thus, despite the fact that the muon flux and showers are generally directed from top to bottom, the intensity of isolated $ev$-neutrons escaping from the chamber walls is
approximately the same over the entire chamber surface. The attenuation length of ev-neutrons averaged over the rock composition and density ($A$  = 22, $Z$  = 11,
${\rho}$ = 2.65 g cm$^{-3}$) is ${\lambda}_n^{is}$ $\approx$ 25 g cm$^{-2}$ \cite{23}, which determines the small thickness of the rock layer, about
0.3 m, from which the $ev$-neutrons escape into the chamber. The energy spectrum of isolated $ev$-neutrons is a composition of the Maxwellian spectra
$F^M(T_n)$ $\propto$ $T_n exp(T_n/T^0)$ with different nuclear temperatures
$T^0$ related to the neutron generation mechanisms in $h$- and $em$-showers and the spectrum of $\delta$-neutrons $F^s(T_{\delta})$ $\propto$ $1/T_{\delta}$  at $T_{\delta} <$ 30 MeV. The mean
energy of cosmogenic $ev$-neutrons is 4 $-$ 7 MeV \cite{2}, \cite{24};
at the exit from the rock their energy decreases to 1 $-$ 3 MeV mainly due to their quasi-elastic scattering by rock nuclei. The low energy of cosmogenic $ev$-neutrons
allows the background produced by them to be efficiently suppressed using a detector shield that incorporates active and passive elements.

As has already been noted, cg-neutrons with energies $T_n >$  30 MeV, i.e., $\delta$-neutrons, have the greatest penetrating power. The flux of $\delta$-neutrons underground is characterized by a significant directivity from top to bottom. This is explained, first, by the strong anisotropy of the angular distribution of muons
$I_{\mu}(\theta)$ $\propto$ $cos^n{\theta}_{\mu}$ with an exponent $n \approx$  2 $-$ 5 at a depth of 1000 $-$ 4000 m w.e. under the flat surface, second, by
the development of an $h$-shower along the muon trajectory, and, third, by the escape of $\delta$-neutrons from nuclei in a narrow cone along the motion of the shower pions. The energy dependence of the total $nA$-interaction cross section in the range 30 $-$ 150 MeV in a standard rock is similar to the energy dependence of the cross section ${\sigma}_{nN}$ $\propto$ $T_n^{-1}$ (Fig. 8). The mean free
path of $\delta$-neutrons ${\lambda}_{\delta}$ with energies 30 $-$ 150 MeV lies within the range from 8 to 40 cm. The above peculiarities of cosmogenic particles underground in combination with the $\delta$-neutron isolation condition determine
the region from which the neutrons arrive at the detector. It is located above the detector and around its upper part. Apart from the relationship between the detector and chamber surfaces, the shape and sizes of the region depend on the mean free path ${\lambda}_{\delta}$ and the angular distribution of $\delta$-neutrons. The number $P_D(T_{\delta})$ of $is$-neutrons incident on the detector is proportional
to the volume of this region $V(T_{\delta})$: $P_D(T_{\delta})$ $\propto$  $V(T_{\delta})$. The longitudinal and transverse sizes of the region are
related to the mean free path ${\lambda}_{\delta}$ and the angular distribution
of $\delta$-neutrons, respectively. In that case, the volume $V(T_{\delta})$ can be represented as a cylinder with a height proportional to ${\lambda}_{\delta}$ and a cross section proportional to the area ${\pi}R_{\delta}^2({\theta}_{\delta})$:
\[
V(T_{\delta}) \propto {\lambda}_{\delta}(T_{\delta})R_{\delta}^2({\theta}_{\delta}).
\]

The radius $R_{\delta}$ is related to the ratio 
\[
p_F^{\perp}/p_r = tan{\theta}_{\delta} {\propto} R_{\delta},
\]
where $p_F^{\perp}$ is the mean transverse Fermi momentum of the nucleon in the nucleus, $p_r$ is the momentum of the $r$-neutron, ${\theta}_{\delta}$ is the escape angle of the released $\delta$-neutron from the nucleus relative to the momentum $p_r$ directed along the motion of the shower pion that is slightly deflected from the muon trajectory and the $h$-shower axis.

Thus, the spectrum of $\delta$-neutrons in the source $F^s(T_{\delta})$ under the action of the factor $V(T_{\delta})$ is transformed into the spectrum of detected isolated neutrons $F^{is}(T_{\delta})$:

\begin{equation}
F^{is}(T_{\delta}) \propto V(T_{\delta})F^s(T_{\delta}) \propto [{\lambda}_{\delta}(T_{\delta})tan^2{\theta}_{\delta}]F^s(T_{\delta}).
\end{equation}

In the range 30 $-$ 150 MeV the mean free path ${\lambda}_{\delta}$ $\propto$ $T_{\delta}$, because ${\sigma}_{nA} \propto$ $1/T_{\delta}$. Since at these energies $F^s(T_{\delta})$ $\propto$ $1/T_{\delta}$, $F^{is}(T_{\delta})$ $\propto$ $tan^2{\theta}_{\delta}$ = $(p_F^{\perp}/p_r)^2$. The dependence of $tan{\theta}_{\delta}$ on $T_{\delta}$  obtained at $p_F^{\perp}c$ = 239 MeV ($T_F$ = 30 MeV) and $T_r  = T_{\delta}$ + 40 MeV is presented in Fig. 9. When normalized at $T_{\delta}$ = 30 MeV ($T_r$ = 70 MeV, $p_rc$ = 369 MeV), it is well fitted by the function $tan{\theta}_{\delta}$ = 1.8$T_{\delta}^{-0.3}$. Consequently, in the range of isolated $\delta$-neutron energies $T_{\delta}^{is}$ from 30 MeV to the knee ($\sim$ 100 MeV) the spectrum $F^{is}(T_{\delta})$ underground is

\begin{equation}
F^{is}(T_{\delta}) \propto tan^2{\theta}_{\delta} \propto T_{\delta}^{-0.6}.
\end{equation}

 \begin{figure}[!t] 
  \centering 
  \includegraphics[width=\linewidth]{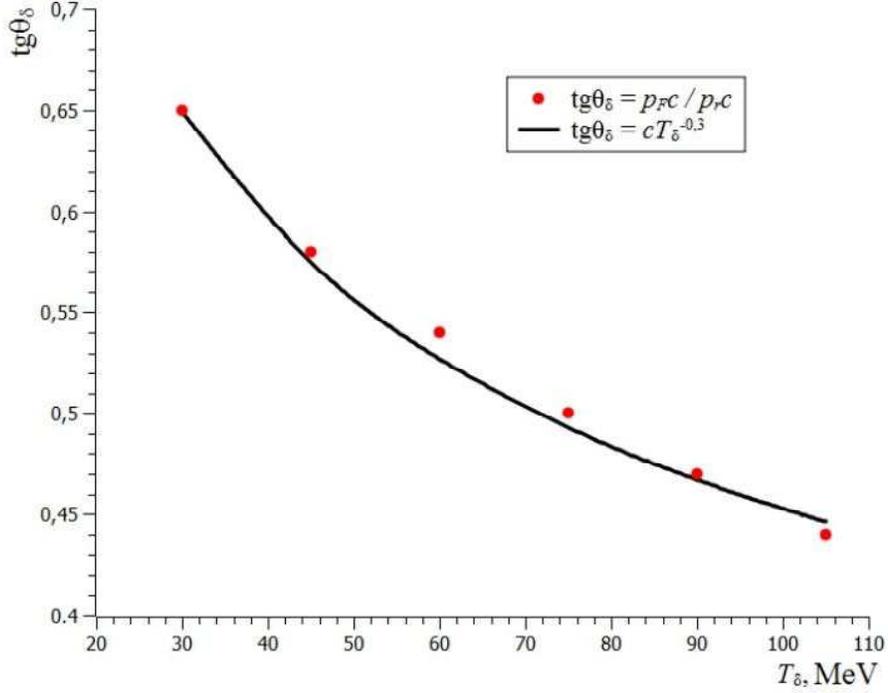} 
  \caption{Relationship between the dependence $tan{\theta}_{\delta}$ = $p_F^{\perp}/p_r$ and the function $T_{\delta}^{-0.3}$ normalized to
$tan{\theta}_{\delta}$ at $T_{\delta}$ =  30 MeV.} 
  \label{9fig} 
 \end{figure}

The shape of the spectrum $F^{is}(T_{\delta})$ in the range $T_{\delta}^{cr}$ $\leq$ $T_{\delta}$ $\leq$ $T_{\delta}^{max}$ $\approx$ 1000 MeV is determined by the quantity ${\lambda}_{\delta}(T_{\delta})$ and the dependence of $tan{\theta}_{\delta}$ on $T_{\delta}$. If we
assume that in the range 150 $-$ 1000 MeV ${\lambda}_{\delta} \approx$ const (Fig. 8) and $tan^2{\theta}_{\delta}$ $\propto$ $T_{\delta}^{-0.6}$, then at $T_{\delta} \approx T_{\delta}^{cr}$ the spectrum $F^{is}(T_{\delta})$ steepens sharply:

\begin{equation}
F^{is}(T_{\delta}) \propto {\lambda}_{\delta}(tan^2{\theta}_{\delta})F^s(T_{\delta}) \propto T_{\delta}^{-0.6}T_{\delta}^{-2} = T_{\delta}^{-2.6}.
\end{equation}

The spectrum $F^{is}(T_{\delta})$ in the energy range 20 $-$ 90 MeV
was measured with the 100-ton cylindrical scintillation detector with approximately equal height and diameter of about 5.5 m. The detector is in a salt mine at a depth of 570 m w.e. \cite{6}. The energy range was determined by the capabilities of the electronics oriented to detect the ${\overline {\nu}}_ep$ reaction caused by neutrinos from collapsing stellar cores. Neutrons were generated in the salt by a muon flux with a mean energy of $\sim$ 125 GeV.
The neutron energy was determined from the energy release of recoil protons in the elastic $np$-scattering reaction and $n$C-interaction particles taking into account quenching in a liquid hydrocarbon scintillator. The spectral index was determined with a 20\% accuracy (Fig. 10): $F^{is}(T_{\delta})$ $\propto$ $T_n^{-0.5 \pm 0.1}$. As can be seen from Fig. 10, the spectrum \cite{6} also agrees satisfactorily
with the dependence $T_{\delta}^{-0.6}$. The spectrum of isolated neutrons generated in $h$-showers in the rock at a depth of 60 m w.e. by muons at a mean energy of $\sim$ 20 GeV was measured in \cite{5} in the range 10 $-$ 60 MeV (Fig. 10). The neutron energy was also determined from the energy release of recoil protons in a 3.85-kg cylindrical plastic scintillator with a dimeter and height of 17.5 cm. The spectrum \cite{5} in the range from $\sim$ 25 to 40 MeV is slightly steeper than the spectrum \cite{6}. The fall of the spectrum at $T_n >$ 40 MeV is explained by the edge effect, which leads to an underestimation of the recoil proton energy. Given the measurement uncertainty, it can be concluded that, being consistent with the experimentally best results of measurements \cite{6}, the dependence (8) explains the shape of the spectrum of isolated $\delta$-neutrons with energies from 30 to $\sim$ 100 MeV produced in the rock. Furthermore, the dependence (9) predicts the behavior of the spectrum $F^{is}(T_{\delta})$ in the range of energies above $\sim$ 100 MeV, which is not yet provided with measurements.

 \begin{figure}[!t] 
  \centering 
  \includegraphics[width=\linewidth]{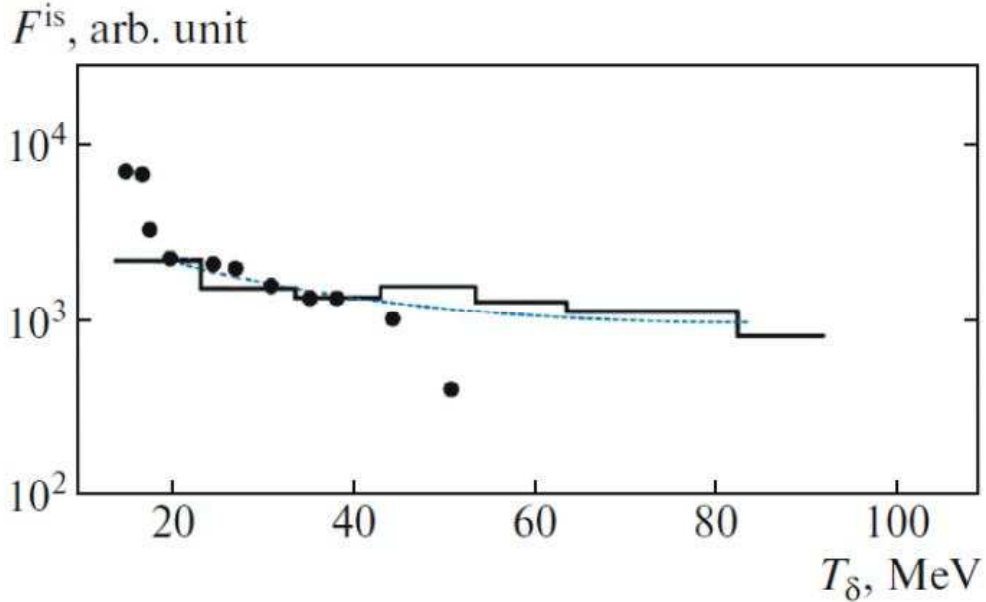} 
  \caption{Spectra of $is$-neutrons. The dots are the results from \cite{5}, the histogram represents the data from \cite{6}, the dashed line indicates the dependence $T_{\delta}^{-0.6}$. All data were normalized to the value of the histogram at $T_{\delta}$ = 20 MeV.} 
  \label{10fig} 
 \end{figure}

\section{The reaction of $\delta$-neutrons in the total number of cg-neutrons}

It is well known from experiments that the number of $\delta$-neutrons $N_{\delta}$ with energies $T_{\delta} \geq$  30 MeV is approximately
half the number of evaporative neutrons $N_h^{ev}$ produced in $h$-showers \cite{2}. Furthermore, it was found in calculations \cite{11},\cite{25}, \cite{26} that the $em$-shower neutrons $N_{em}$ account for 25 $-$ 40\% of the total number of cosmogenic neutrons $N_n^{tot}$, depending on the material.
Assuming that all neutrons $N_{em}$ belong to the evaporative region, we arrive at the equation

\begin{equation}
N_n^{tot} = N_{\delta} + N_{h}^{ev} +N_{em}\\
= N_{\delta} + 2N_{\delta} + N_{em} = 3N_{\delta} +{\eta}N_n^{tot}
\end{equation}

Hence it follows that $N_{\delta}  = (1 - {\eta})N_n^{tot} /3$  and at
$\eta$ = 0.25 $-$ 0.40
\begin{equation}
N_{\delta} = (0.20 - 0.25)N_n^{tot}
\end{equation}
Consequently, the number of $\delta$-neutrons at energies 30 $-$ 1000 MeV and evaporative neutrons (with energies up to 30 MeV) accounts for, 
respectively, 20 $-$ 25\% and 75 $-$ 80\% of $N_n^{tot}$.

\section{Conclusions}

The cg-neutron production spectrum is explained in terms of the quark model of deep-inelastic ${\pi}A$-scattering. The spectrum is formed by the interactions in nuclei of atoms of the medium. The characteristics of the spectrum (the shape, the mean energy, the knee energy, the mean energies of the spectral ranges) are related to the mechanism for the $qq$ collision of the valance quark of an incident pion with the valence quark of a nucleon in the target nucleus and the passage of the recoil neutron through the nuclear matter.

The spectrum of cg-neutrons occupies the range from zero to $\sim$ 1 GeV and consists of three components. The first component (energies up to 30 MeV) is
dominated by evaporative neutrons with a Maxwellian spectrum. The $\delta$-neutrons with a spectrum $1/T_{\delta}$ present here have virtually no effect on the shape of the resulting spectrum. The mean energy 4 $-$ 7 MeV of neutrons in this spectral range is determined by the value of $A$ and is unrelated to ${\overline E}_{\mu}$. The neutrons of thefirst component account for 75 $-$ 80\% of the total number of cg-neutrons, depending on $A$. The $\delta$-neutrons of deep-inelastic ${\pi}A$-scattering constitute the second and third components spanning the energy range from 30 MeV to $\sim$ 1 GeV. The spectrum of $\delta$-neutrons $F^s(T_{\delta})$ has a characteristic shape representable in a logarithmic
scale by a straight line with a knee at energy $T_{\delta}^{cr}$.

The source of $\delta$-neutrons with energies above 30 MeV are $h$-showers. The neutrons of $em$-showers and those produced in other processes do not affect
the shape of the spectrum. The shape of the spectrum $F^s(T_{\delta})$ originates from the energy spectrum of $\delta$-particles $1/T_{\delta}^2$ that characterizes the elastic scattering of quasi-free valance quarks. The spectrum $F^s(T_{\delta})$ is formed by those of the recoil neutrons that escape from the nuclei without collisions. As a result of the interactions of recoil neutrons in the nuclei, the spectrum $1/T_{\delta}^2$ in the range from 30 MeV to the knee energy $
T_{\delta}^{cr}$ takes the shape of the second component $1/T_{\delta}$. The neutrons of the third component from $T_{\delta}^{cr}$ to $\sim$ 1 GeV 
have a spectrum $1/ T_{\delta}^2$, because they all escape from the nuclei without collisions. The energy of the knee $T_{\delta}^{cr}$ in the 
spectrum $F^s(T_{\delta})$ depends on $A$ as $T_{\delta}^{cr}$ $\propto$ $A^{1/4}$; for nuclei with $A \leq 4$ the spectrum $F^s(T_{\delta})$ takes 
a mono shape proportional to $T_{\delta}^{-2}$, while for $A$ = 207 (Pb) the second component of the spectrum proportional to $T_{\delta}^{-1}$ 
reaches an energy of $\sim$ 190 MeV.

The mean energy of $\delta$-neutrons, ${\overline T}_{\delta}$ $\sim$ (120 $-$ 150) MeV is weakly related to $A$ and does not depend on the shower pion energy. As a consequence, the shape of the spectrum $F^s(T_{\delta})$ and the energy ${\overline T}_{\delta}$ of neutrons generated in the rock remain unchanged with increasing depth and correspondingly rising mean muon energy ${\overline E}_{\mu}$. For the same reason, the shape of the spectrum
$F^s(T_{\delta})$ in the atmosphere must closely follow the shape of the cg-neutron spectrum underground.

The energy spectrum of isolated neutrons $F^{is}(T)_{\delta}$ also has a three-component shape. Here, the mean energy of $ev$-neutrons of the first component is lower than the mean energy of $ev$-neutrons of the production spectrum due to the scattering of neutrons by nuclei in the medium. The spectrum of $is$-neutrons
underground $F^{is}(T_{\delta})$ at energies above 30 MeV is formed by the $\delta$-neutrons arriving at the detector from the region of their generation without noticable change in energy and direction. The location of this area relative to the detector and its dimensions are determined by the angular distribution of muons at the depth of observation, the angular distribution of $\delta$-neutrons as they escape from the nuclei, and the mean free path in the rock dependent on their energy. The main
factor that determines the shape of the spectrum $F^{is}(T_{\delta})$ is the dimensions of the area from which the $\delta$-neutrons reach the detector. Under the action of this factor the second component of the spectrum $F^s(T_{\delta})$ acquires the shape $F^{is}(T_{\delta})$ $\propto$ $T_{\delta}^{-0.6}$, while the third
component proportional to $T_{\delta}^{-2}$ is transformed into $T_{\delta}^{-2.6}$.

One of the revealed important peculiarities is the cutoff of the spectrum at an energy of $\sim$ 1 GeV corresponding to a maximum transverse momentum $p_{\perp}$ $\sim$  2 GeV/$c$ in deep-inelastic soft processes. No cutoff has been established so far in measurements, because it is fairly difficult to determine the energy of cg-neutrons in this energy range experimentally. The spectra of cg-neutrons obtained by the Monte Carlo simulation show no sharp change in their shape at energies above 1 GeV
(Fig. 1). At the same time, the effect of limiting nuclear fragmentation in a deep-inelastic hadron– nucleus interaction and the behavior of the cross section ${\sigma}_{nN}$
 (Fig. 3) at $T_n \geq$ 1 GeV (implying the dominance of pion production in ${\pi}N$-collisions with neutron energy loss both inside and outside the nucleus) point to the existence of a cutoff in the spectrum of cg-neutrons.

The Monte Carlo simulations of the cg-neutron spectrum available to date are inconsistent between themselves and with the measurements. The proposed
fits to the data of these simulations by the function $T_n^{-\alpha}$
at $\alpha$ = const \cite{27}, the sum of exponentials $exp(-aT_n)$,
or a combination of these functions \cite{12}, \cite{28}, \cite{29} do not
clarify the cg-neutron spectrum formation mechanism.
Monte Carlo calculations do not yet provide simulations of the generation of neutrons by highenergy muons that fit the real one. Knowing the physical processes shaping the spectrum of cg-neutrons allows the Monte Carlo codes needed for the optimization of experiments and the calculation of background effects to be selected and corrected.
~~~

{\bf Acknowledgements.} 
This work was supported by the Russian Foundation
for Basic Research (project No. 15-02-01056a, 18-02-00064a)
and the program of investigations of the Presidium
of Russian Academy of Sciences High Energy Physics
and Neutrino Astrophysics.

\end{document}